\documentstyle[twoside,fleqn,espcrc2,epsf]{article}
\renewcommand{\floatsep}{5mm plus 2pt minus 0pt}
\textfloatsep=\floatsep
\intextsep=\floatsep
\newcommand{\Journal}[4]{{#1} {\bf #2}, #3 (#4)}

\newcommand{\NPB}{Nucl.~Phys.~B}
\newcommand{\PLB}{Phys.~Lett.~B}
\newcommand{\PRL}{Phys.\ Rev.~Lett.\ }

\newcommand{\Sp}{{\rm sp}}
\newcommand{\tp}{{\rm tp}}
\newcommand{\sr}{{\rm sr}}
\newcommand{\ttr}{{\rm ttr}}
\newcommand{\str}{{\rm str}}

\title{\bf Improved gluonic actions on anisotropic lattices}

\author{Colin Morningstar\address{Dept.~of Physics,
  University of California at San Diego,
  La Jolla, California 92093-0319}}

\begin{document}

\begin{abstract}
The use of novel perturbatively-improved gluonic actions
on anisotropic lattices in which the temporal spacing is much
smaller than that in the spatial directions is discussed.
Such actions permit more efficient measurements of noisy
correlation functions, such as glueball correlators, on
coarse lattices.  A derivation of these actions at tree-level
is outlined; mean-field link renormalization plays a crucial
role in their construction.  Results for the low-lying glueball
masses and the heavy-quark potential are presented.
\end{abstract}

\maketitle

The use of improved actions makes possible accurate Monte Carlo
simulations of QCD on coarse lattices with greatly reduced
computational effort.  However, for some calculations, such
as glueball masses\cite{morningstar95}, the coarseness of the
temporal lattice spacing can be a severe drawback, greatly reducing
the number of correlator time intervals which can be measured.
This problem can be circumvented by using anisotropic lattices
in which the temporal spacing is much smaller than that in the
spatial directions, enabling one to exploit the enhanced
signal-to-noise of the correlation functions at smaller
temporal separations.  Improved lattice actions often admit
spurious states with energies of order $1/a$, where $a$ is the
lattice spacing.  On fine grids, such modes are comfortably
high-lying and little affect the low-lying modes of interest.
However, on a coarse lattice, this may no longer be true.  Reducing
the temporal spacing also cures this problem by elevating the
spurious states to higher energies of order $1/a_t$, where
$a_t$ is the temporal lattice spacing.  The lifting of spurious
states is particularly important when designing improved fermion
actions\cite{Klassen}.  In fact, anisotropic
lattices will be useful whenever one is faced with a four-momentum
in which one component is unusually large, such as in the
calculation of glueball masses\cite{peardon}, the heavy quarkonium
spectrum, or hadronic form factors at large momentum transfers.
Anisotropic lattices have also long been used in finite
temperature studies.

To remove $O(a^2)$ errors from all Green's functions
in lattice QCD at tree level in perturbation theory, one needs
only to eliminate $O(a^2)$ artifacts in the lattice action.
This can be done by performing a small-$a$ expansion
of some suitable lattice action and adjusting the interaction
couplings so that the leading terms in the expansion reproduce the
correct continuum QCD action and the $O(a^2)$ terms are absent.
Any lattice operator $W_\beta[U]$ constructed from link variables
will have a small-$a$ expansion of the form
\begin{equation}
W_\beta[U] = \sum_{k=0}^\infty \int\!\! d^4x\,
  \xi^{-1}\! a_s^{k-4}\!\! \sum_{\alpha=1}^{M(k)}
   c^{(k)}_\alpha\,  Q^{(k)}_\alpha(x),
\label{smalla}
\end{equation}
where $c^{(k)}_\alpha$ are the expansion coefficients, the
spatial lattice spacing is $a_s$, the
temporal spacing is $a_t=\xi a_s$, and
$Q^{(k)}_\alpha(x)$ are local, dimension-$k$ continuum operators
at point $x$ which are invariant under gauge transformations
and all symmetries of the anisotropic lattice.  There are no such operators
for dimensions less than four.  There are only two
dimension-four operators:
$Q^{(4)}_1 = g^2 {\rm Tr} {\bf E}^2$ and
$Q^{(4)}_2 = g^2 {\rm Tr} {\bf B}^2$,
where ${\bf E}$ and ${\bf B}$ are the chromoelectric and
chromomagnetic fields, respectively.  There are no
dimension-five operators, but there are 18 independent
dimension-six operators, ten of which may be expressed as
unimportant total derivatives.  Thus,
we adjust the couplings of the $W_\beta[U]$ lattice operators
so that the coefficients of the eight dimension-six operators
vanish and the coefficients of the two dimension-four operators
equal each other.  If we are interested only in {\em on-shell}
improvement, we are also free to use a field
redefinition to set some of the coefficients of the
dimension-six operators to zero.

Perturbation theory by itself does not reliably determine the
couplings in the improved action.  It is known that a judicious
combination of perturbation theory and mean field theory works
much better.  Mean field theory is introduced
by renormalizing the link variables:
$U_j(x) \rightarrow U_j(x) / u_s$ and
$U_t(x) \rightarrow U_t(x) / u_t$,
where $u_t$ and $u_s$ are the mean values of the gluon link
operators for the temporal and spatial links, respectively.

Carrying out the above procedure, one of the simplest actions
we can arrive at is given by\cite{aniso}:
\begin{eqnarray}
S_I[U]& = &\beta\xi\ \Bigl\{ \frac{5}{3u_s^4}W_\Sp
    +\frac{5}{3\xi^2u_s^2u_t^2} W_\tp  \nonumber\\
     &-& \frac{1}{12u_s^6} W_\sr
     - \frac{1}{12\xi^2u_s^4u_t^2} W_\str \nonumber\\
     &-& \frac{1}{12\xi^2u_s^2u_t^4} W_\ttr \Bigr\},
\label{actiond}
\end{eqnarray}
where $\beta\!=\!6/g^2$, $W_c=\sum_c {\textstyle\frac{1}{3}}
\mbox{Re Tr}(1-U_c)$, and
$U_\Sp$ denotes the spatial plaquettes, $U_\tp$ indicates
the temporal plaquettes, $U_\sr$ denotes the product of link
variables about a planar $2\times 1$ spatial rectangular loop,
$U_\str$ refers to the short temporal rectangles (one temporal
spacing, two spatial), and $U_\ttr$ refers to the
tall temporal rectangles (two temporal spacings, one spatial).
The mean values $u_t$ and $u_s$ are determined by guessing input
values for use in the action, measuring the mean links in a
simulation, then readjusting the input values accordingly.
When $a_t$ is significantly
smaller than $a_s$, we expect the mean temporal link $u_t$ to be
very close to unity.  For example, in Landau-gauge perturbation theory,
$1-\langle\mbox{$\frac{1}{3}$}{\rm Tr} U_t\rangle \propto (a_t/a_s)^2$.
Hence, to simplify matters, we set $u_t\!=\!1$.   A convenient and
gauge-invariant definition for $u_s$ in terms of the mean spatial
plaquette is then given by
$u_s=\langle \frac{1}{3}{\rm ReTr} P_{ss^\prime}\rangle^{1/4}$,
where $P_{ss^\prime}$ denotes the spatial plaquette.

\begin{table}[t]
\caption[tabone]{Renormalization of the anisotropy.  The input
anisotropy is $\xi=a_t/a_s$, and $\xi_{\rm meas}$ is the
measured anisotropy (see text).  The lattice spacing $a_s$ is set
using the $1P-1S$ splitting in charmonium for $SU_3$ and the string
tension for $SU_2$. $S_{LW}$ is the improved L\"uscher-Weisz
action\protect\cite{LW}.}
\label{renaniso}
\setlength{\tabcolsep}{1.5mm}
\begin{tabular}{cccll}
\hline
Action & $\beta$ & $\xi$ & $\xi_{\rm meas}$ & $a_s$ (fm) \\ \hline
$S_{LW}$        &  $1.9$  &  $1/1$  &  $1.00$      &  $0.397(4)$  \\
$S_{II}$        &  $2.0$  &  $1/2$  &  $0.505(11)$ &  $0.418(6)$  \\
                &  $2.0$  &  $1/3$  &  $0.342(10)$ &  $0.372(6)$  \\
                &  $2.3$  &  $1/2$  &  $0.507(10)$ &  $0.311(5)$  \\
$S_I$           &  $2.0$  &  $1/2$  &  $0.505(10)$ &  $0.371(4)$  \\
                &  $2.3$  &  $1/2$  &  $0.510(11)$ &  $0.271(3)$  \\
$S_{II}(u_s\!\!=\!\!1)$
                &  $3.9$  &  $1/2$  &  $0.40(2)$   &  $0.40(2)$   \\
$S_{II}(SU_2)$  & $0.848$ & $0.276$ &  $0.286(6)$  &  $0.371(9)$  \\
                & $1.027$ & $0.351$ &  $0.357(4)$  &  $0.283(6)$  \\
                & $1.114$ & $0.409$ &  $0.417(6)$  &  $0.249(6)$  \\
              \hline
\end{tabular}
\end{table}

One defect of the action $S_I$ is that the gluon spectrum has
spurious high-energy states arising from $W_{\ttr}$, which spans
two time slices.  These modes occur at energies of order $2/a_t$
and have little effect on simulation results, but they can cause
problems when applying the variational method to extract masses
from short-time correlation functions.  These modes may be
eliminated by relaxing the improvement conditions,
requiring that the coefficients of all dimension-six operators
{\em except} ${\rm Tr} [(D_t{\bf E})^2]$ vanish.  One then
obtains:
\begin{eqnarray}
S_{II}[U]& = &\beta\xi\ \Bigl\{ \frac{5}{3u_s^4}W_\Sp
    +\frac{4}{3\xi^2u_s^2u_t^2} W_\tp  \nonumber\\
     &-& \frac{1}{12u_s^6} W_\sr
     - \frac{1}{12\xi^2u_s^4u_t^2} W_\str \Bigr\}.
\label{action}
\end{eqnarray}
This action has $O(a_t^2)$ errors which are very small,
being suppressed by a factor of $\xi^4$.  For this action,
various values for the parameter $u_s(\beta,\xi)$ are
$u_s(1.7,\frac{1}{3})=0.745$,
$u_s(2.0,\frac{1}{3})=0.772$, and
$u_s(2.4,\frac{1}{3})=0.806$.

%\section{Renormalized Anisotropy}

\begin{figure}[t]
\begin{center}
\leavevmode
\epsfxsize=3in\epsfbox[80 200 530 465]{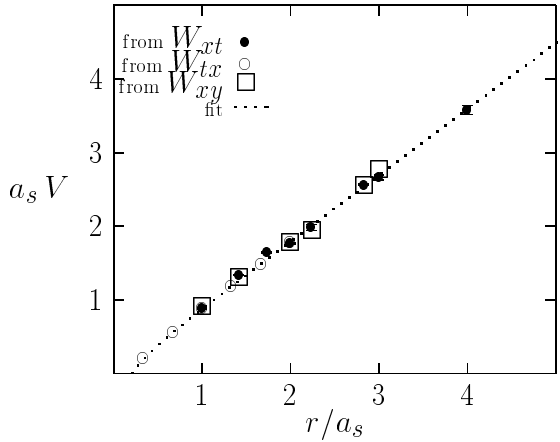}
\end{center}
\caption{The static-quark potential from Wilson loops
in different orientations. Results are shown for the action $S_{II}$
with $\beta=2.0$ and $a_t/a_s=1/3$.  The different
potentials were shifted to agree at~$r\!=\!a_s$.}
\label{vfig}
\end{figure}

The renormalization of the anisotropy can be determined by measuring
the static-quark potential $V(x,y,z)$ from Wilson loops in different
orientations.  For example, Wilson loops in the $xt$ and $xy$ hyperplanes
have the following asymptotic behaviours ($I,J$ positive integers):
\begin{eqnarray}
W_{xt}(Ia_s,Ja_t)\! &\stackrel{J\rightarrow\infty}{\longrightarrow}&
\! Z_{xt} e^{-Ja_t V(Ia_s,0,0)},\\
 W_{xy}(Ia_s,Ja_s)\! &\stackrel{J\rightarrow\infty}{\longrightarrow}&
\! Z_{xy} e^{-Ja_s [V(Ia_s,0,0)+V_0]},
\end{eqnarray}
where $Z_{xt}$, $Z_{xy}$, and $V_0$ are renormalization constants.
{}From the asymptotic behaviour of $W_{xt}$ we obtain
$\Delta_{xt}\!=\!a_t[V(2a_s,0,0)\!-\!V(a_s,0,0)]$
and from $W_{xy}$ we obtain $\Delta_{xy}\!=\!a_s[V(2a_s,0,0)
\!-\!V(a_s,0,0)]$.  The renormalized anisotropy can then be defined
as the ratio $\xi_{\rm meas}=\Delta_{xt}/\Delta_{xy}$.
Results for the renormalized anisotropy\cite{aniso}
are listed in Table~\ref{renaniso}.
Note that without tadpole improvement $(u_s\!=\!1)$, the
renormalization of the anisotropy is a $20\%$ effect; mean-field
improvement reduces this renormalization to the $1-3\%$ level.
Results for the static-quark potential\cite{aniso}
are shown in Fig.~\ref{vfig}.

\begin{figure}[t]
\begin{center}
\leavevmode
\epsfxsize=3in\epsfbox[80 100 530 490]{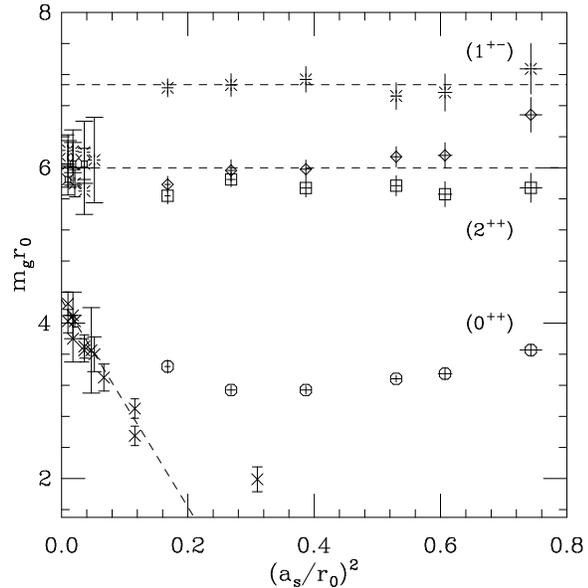}
\end{center}
\caption[glue]{Low-lying glueball masses $m_g$ against
lattice spacing $a_s$ in terms of the hadronic scale
$r_0\approx 1/2$ fm defined by $[r^2 dV(r)/dr]_{r=r_0}=1.65$
where $V(r)$ is the static quark potential.
The $\times$ denote results for the scalar and tensor glueballs
using the simple Wilson action\protect\cite{WG}.  Results
using the improved action $S_{II}$
for the $A_1^{++}$, $E^{++}$, $T_2^{++}$,
and $T_1^{+-}$ representations are indicated by
$\circ$, $\Box$, $\Diamond$, and $\ast$, respectively.
The dashed lines are linear fits to the Wilson results for the
scalar and tensor; for the $1^{+-}$, the dashed line guides the eye.
\label{fig:scaling}}
\end{figure}

The masses of various low-lying
glueballs were also calculated\cite{peardon}.
The results are shown in Fig.~\ref{fig:scaling}
as a function of the lattice spacing as measured in terms of
the hadronic scale $r_0$.  The scalar glueball mass
from the improved action $S_{II}$ exhibits dramatically
reduced cutoff contamination compared to the Wilson action.
Finite-$a_s$ errors are seen to be small for the tensor and
pseudovector glueballs, although differences between
the $E^{++}$ and $T_2^{++}$ representations indicate
violations of rotational invariance, especially
for large $a_s$.

\epsfverbosetrue

\epsfverbosetrue

These results clearly demonstrate the usefulness of
perturbatively-improved gluonic actions on anisotropic lattices.
For coarse-lattice glueball studies, anisotropic lattices are crucial.
The effectiveness of tadpole improvement in reducing the
renormalization of the anisotropy to the level of a few per cent
makes using anisotropic lattices no more difficult than isotropic ones.
This work was supported by grants from the NSF, the DOE, and the
NSERC of Canada.


\begin{thebibliography}{9}
\bibitem{morningstar95}
 C.~Morningstar and M.~Peardon, Nucl.\ Phys. B (Proc.\ Suppl.)
 {\bf 47}, 258 (1996).
\bibitem{Klassen}
 M.~Alford and T.~Klassen, these proceedings.
\bibitem{peardon}
 C.~Morningstar and M.~Peardon, these proceedings.
%\bibitem{TI} G.P.~Lepage and P.~Mackenzie,
%  \Journal{\PRD}{48}{2250}{1993}.
\bibitem{aniso}
 M.~Alford, T.~Klassen, P.~Lepage,
 C.\ Morningstar, M.~Peardon, H.~Trottier,
 to appear.
\bibitem{LW}
 M.~L\"uscher and P.~Weisz, Comm.\ Math.\ Phys.\ {\bf 97}, 59 (1985).
\bibitem{WG} J.~Sexton, {\em et al.},
  \Journal{\PRL}{75}{4563}{1995}; P.~De Forcrand {\em et al.},
  \Journal{\PLB}{152}{107}{1985}; C.~Michael and M.~Teper,
  \Journal{\NPB}{314}{347}{1989}; UKQCD Collaboration,
  \Journal{\PLB}{309}{378}{1993}.
\end{thebibliography}
\end{document}